\documentclass{emulateapj}

\usepackage{graphicx,amssymb,amsmath,times}

\begin{document}
%Title of paper
\title{Observation of Gamma Rays from the Galactic Center with the MAGIC Telescope}

\author{
 J.~Albert\altaffilmark{a}, 
 E.~Aliu\altaffilmark{b}, 
 H.~Anderhub\altaffilmark{c}, 
 P.~Antoranz\altaffilmark{d}, 
 A.~Armada\altaffilmark{b}, 
 M.~Asensio\altaffilmark{d}, 
 C.~Baixeras\altaffilmark{e}, 
 J.~A.~Barrio\altaffilmark{d}, 
 M.~Bartelt\altaffilmark{f}, 
 H.~Bartko\altaffilmark{g,*}
% ...
}
 \altaffiltext{a} {Universit\"at W\"urzburg, D-97074 W\"urzburg, Germany}
 \altaffiltext{b} {Institut de F\'\i sica d'Altes Energies, Edifici Cn., E-08193 Bellaterra (Barcelona), Spain}
 \altaffiltext{c} {ETH Z\"urich, CH-8093 H\"onggerberg, Switzerland}
 \altaffiltext{d} {Universidad Complutense, E-28040 Madrid, Spain}
 \altaffiltext{e} {Universitat Aut\`onoma de Barcelona, E-08193 Bellaterra, Spain}
 \altaffiltext{f} {Universit\"at Dortmund, D-44227 Dortmund, Germany}
 \altaffiltext{g} {Max-Planck-Institut f\"ur Physik, D-80805 M\"unchen, Germany}
% ...
\altaffiltext{*} {correspondence: hbartko@mppmu.mpg.de}

%\enlargethispage{0.5cm}

\begin{abstract}

Recently, the Galactic Center has been reported to be a source of very high energy (VHE) $\gamma$-rays by the VERITAS, CANGAROO and HESS experiments. The energy spectra as measured by these experiments show substantial differences. In this {\it Letter} we present MAGIC observations of the Galactic Center, resulting in the detection
of a differential $\gamma$-ray flux consistent with a steady, hard-slope
power law, described as $\mathrm{d}N_{\gamma}/(\mathrm{d}A \mathrm{d}t \mathrm{d}E) = (2.9 \pm 0.6)\times 10^{-12} (E/\mathrm{TeV})^{-2.2 \pm 0.2} \ \mathrm{cm}^{-2}\mathrm{s}^{-1} \mathrm{TeV}^{-1}$. The $\gamma$-ray source is centered at (Ra, Dec)=(17$^{\mathrm{h}}45^{\mathrm{m}}20^{\mathrm{s}}$, -29$^\circ2'$). This result confirms the previous measurements by the HESS experiment and indicates a steady source of TeV $\gamma$-rays. We briefly describe the observational technique used, the procedure implemented for the data analysis, and discuss the results in the perspective of different models proposed for the acceleration of the VHE $\gamma$-rays.

\end{abstract}

\keywords{gamma rays: observation --- Galaxy: center --- acceleration of particles}

\section{Introduction}

The Galactic Center (GC) region contains many remarkable objects which may be 
responsible for high-energy processes generating $\gamma$-rays 
\citep{Aharonian2005,Atoyan2004}. The GC is rich in massive stellar 
clusters with up to 100 OB stars \citep{GC_environment}, immersed in a dense 
gas. There are young supernova remnants, e.g. Sgr A East, and nonthermal radio arcs \citep{LaRosa2000}. The dynamical center of the Milky Way is associated with the compact radio source Sgr A$^*$, which is believed to be a massive black hole \citep{GC_environment}. Within a radius of 10 pc around the GC there is a mass of about $3 \cdot 10^7 M_{\odot}$ \citep{Schoedel2002}. % ,Genzel2003}. 

EGRET has detected a strong source in the direction of the GC, 
3 EG J1745-2852 \citep{GC_egret}, which has a broken power law energy spectrum 
extending up to at least 10 GeV, with a spectral index of 1.3 below the break at a few 
GeV. Assuming a distance of the GC of 8.5 kpc, the $\gamma$-ray luminosity of this source 
is very large, $~2.2 \cdot 10^{37} \mathrm{erg}/\mathrm{s}$, which is 
equivalent to about 10 times the $\gamma$-flux from the Crab nebula. However, an independent analysis of the EGRET data 
\citep{Hooper2002} indicates a point source whose position is different from the GC at a confidence level beyond 99.9 \%. This was recently sustained by \citet{Pohl2005}.

In very high energy $\gamma$-rays the GC has been observed by VERITAS, CANGAROO and HESS \citep{GC_whipple,GC_cangaroo,GC_hess}. The energy spectra as measured by these experiments show substantial differences. This might be due to different sky integration regions of the signal or a source variability at a time-scale of about one year.%, or to inter-calibration problems.

\section{Observations}

MAGIC (see e.g., \citet{MAGIC-commissioning,CortinaICRC} for a
detailed description) is currently the largest single dish Imaging
Air Cherenkov Telescope (IACT) in operation. Located on the Canary
Island La Palma ($28.8^\circ$N, $17.8^\circ$W, 2200~m a.s.l.), the
telescope has a 17-m diameter tessellated parabolic mirror,
supported by a light weight carbon fiber frame. It is equipped
with a high quantum efficiency 576-pixel $3.5^\circ$ field-of-view photomultiplier
camera. The analog signals are transported via optical fibers to
the trigger electronics
and are read out by a 300 MSamples/s FADC system.%the 300 MHz FADC readout.

At La Palma, the GC ((RA, Dec) = $(17^{\mathrm{h}}45^{\mathrm{m}}36^{\mathrm{s}},-28^{\circ}56'))$ culminates at about $58^\circ$ zenith angle (ZA). The star field around the GC is non-uniform. In the region west of the source (RA$\;>\;$RA$_{\mathrm{GC}}+ 4.7^{\mathrm{m}}$) the star field is brighter. Within a distance of 1$^{\circ}$ from the GC there are no stars brighter than 8$^{\mathrm{th}}$ magnitude.

The MAGIC observations were carried out in the ON/OFF mode as well as in the false-source tracking (wobble) mode \citep{wobble}. The sky directions (W1, W2) to be tracked in the wobble mode are chosen such that in the camera the star field relative to the source position (GC) is similar to the star field relative to the mirror source position (anti-source position): W1/W2 = (RA$_{\mathrm{GC}}$, Dec$_{\mathrm{GC}} \pm 0.4^{\circ}$). During one wobble mode data taking, 50\% of the data is taken at W1 and 50\% at W2, switching between the two positions every 20 minutes. Dedicated OFF data have been taken, with a sky field similar to that of the ON region. The OFF region is centered at the Galactic Plane, GC$_{\mathrm{OFF}}$ = (RA, Dec) = $(17^\mathrm{h}51^{\mathrm{m}}12^{\mathrm{s}},\;-26^{\circ}52'00'')$. In the same night OFF data was taken directly before or after the ON observations under the same weather conditions and hardware setup.

\begin{deluxetable}{ccccccc}
\tabletypesize{\scriptsize}
\tablecaption{Data Set}
\tablewidth{\columnwidth} %200pt}
\tablehead{
  \colhead{Period} & \colhead{date} & \colhead{ZA [$^{\circ}$]} & \colhead{time [h]} & \colhead{events [$10^{6}$]} & \colhead{obs. mode} 
}
\startdata
I   & Sep. 2004      & 62-68 & 2     & 0.8     & ON\\
II  & May 2005       & 58-62 & 7     & 2.8     & wobble\\
III & Jun./Jul. 2005 & 58-62 & 17/12 & 6.4/5.0 & ON/OFF\\
\enddata
\tablecomments{Data set per observation period of the GC. The column ``time'' states the effective observation time, the column ``events'' states the events after image cleaning}
\label{tab:observations}  
\end{deluxetable}

After initial observations in September 2004 the GC was observed for a total of about 24 hours in the period May-July 2005. Table \ref{tab:observations} summarizes the data taken.

\section{Data Analysis}

\begin{figure}[!h]
\begin{center}
\includegraphics[totalheight=7cm]{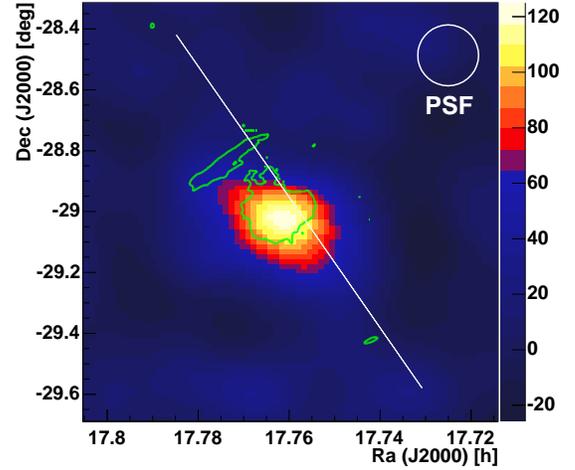}
\end{center}
\caption{Smoothed sky map of $\gamma$-ray candidates (background subtracted) in the direction of the Galactic Center for SIZE $\geq 300$ ph.~el. (corresponding to an energy threshold of about 1~TeV). Overlayed are green contours (0.3 Jy beam$^{-1}$) of 90 cm VLA (BCD configuration) radio data \citep{LaRosa2000}. The white line shows the galactic plane.}
\label{fig:prelresults}
\end{figure}

\begin{figure}[!h]
\begin{center}
\includegraphics[totalheight=6cm]{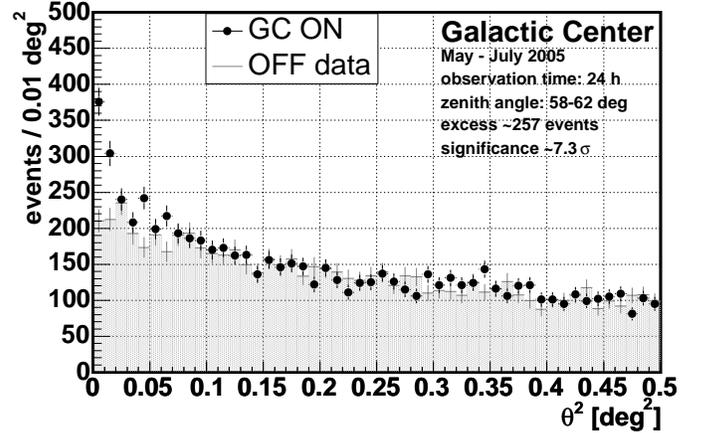}
\end{center}
\caption{Distributions of $\theta^2$ values for the source and anti-source, see text, for SIZE $\geq 300$ ph.~el. (corresponding to an energy threshold of about 1~TeV).}
\label{fig:theta2}
\end{figure}

The data analysis has been carried out using the standard MAGIC analysis and reconstruction software \citep{Magic-software}, the first step of which involves the calibration of the raw data \citep{MAGIC_calibration}. After calibration, image cleaning tail cuts of 10 ph.~el. (core pixels) and 5 ph.~el. (boundary pixels) have been applied (see e.g. \citep{Fegan1997}). These tail cuts are accordingly 
scaled for the larger size of the outer pixels of the MAGIC camera. The camera images are parameterized by image parameters \citep{Hillas_parameters}. 

In this analysis, the Random Forest method (see \citet{RF} for a
detailed description) was applied for the $\gamma$/hadron
separation and the energy estimation (for a review see e.g. \citet{Fegan1997}). For the training of the
Random Forest a sample of Monte Carlo (MC) generated
$\gamma$-showers was used together with about 1\% independent and identically distributed events drawn from the measured OFF-data. The MC
$\gamma$-showers were generated between 58$^\circ$ and 68$^\circ$
ZA with energies between 10~GeV and 30~TeV. For the analysis of the September 2004 data set the Random Forest cuts 
were determined using a sub-set of Galactic OFF data as background.

The source-position independent image parameters  SIZE, WIDTH, LENGTH, CONC \citep{Hillas_parameters} and the third moment of the ph.~el. distribution along the major image axis, as well as the source-position dependent parameter DIST \citep{Hillas_parameters}, were
selected to parameterize the shower images. After the training,
the Random Forest method allows to calculate for every event a
parameter, called hadronness, which is a measure of the
probability that the event is not $\gamma$-like. The
$\gamma$-sample is defined by selecting showers with a hadronness
below a specified value. An independent sample of MC $\gamma$-showers was used to determine the efficiency of the cuts. %applied cuts

\begin{figure}[!h]
\begin{center}
\includegraphics[totalheight=6cm]{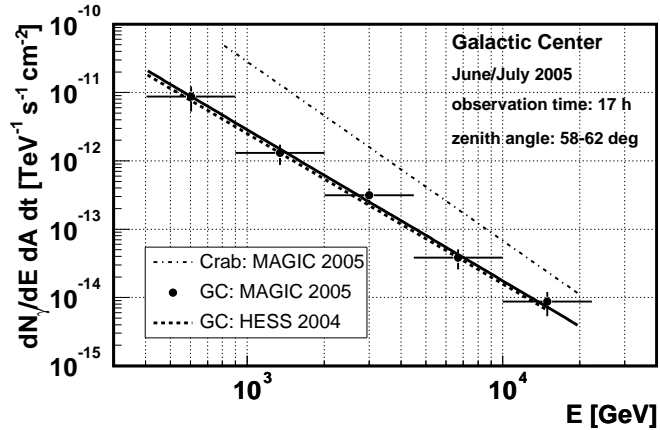}
\end{center}
\caption{Reconstructed VHE $\gamma$-ray energy spectrum of the GC (statistical errors only). The full line shows the result of a power-law fit to the data points. The dashed line shows the result of the HESS collaboration \citep{GC_hess}. The dot-dashed line shows the energy spectrum of the Crab nebula as measured by MAGIC \citep{Crab_MAGIC}.}
\label{fig:spectrum}
\end{figure}

The analysis at high zenith angles was developed and verified using Crab data with a ZA around 60$^{\circ}$. The determined Crab energy spectrum was found to be consistent with other existing measurements (see Fig. \ref{fig:spectrum}, dot-dashed line).

For each event the arrival direction of the primary in sky coordinates is estimated by using the DISP-method \citep{wobble,Lessard2001,MAGIC_disp}. For the sky map calculation only source independent image parameters are used in the Random Forest training. Figure \ref{fig:prelresults} shows the sky map of $\gamma$-ray candidates (background subtracted, see e.g. \citep{Rowell2003}) from the GC region (observation periods II/III). It is folded with a two-dimensional Gaussian with a standard deviation of $0.1^{\circ}$ (roughly corresponding to the MAGIC PSF) and height one. A lower SIZE cut of
300 ph.~el. has been applied, corresponding to an energy
threshold of about 1~TeV. The sky map is overlayed with contours (0.3 Jy beam$^{-1}$) of 90
cm VLA (BCD configuration) radio data from \citet{LaRosa2000}. The brightest non-central source is the Arc. The excess is centered at (RA, Dec)
= (17$^{\mathrm{h}}$45$^{\mathrm{m}}$20$^{\mathrm{s}}$, -29$^\circ2'$) (J2000 coordinates). The systematic pointing uncertainty is estimated to
be $2'$ (for description of the MAGIC telescope drive system see \citet{Bretz2003}) and might in the future be further reduced with the MAGIC starfield
monitor \citep{starguider}. The excess is compatible with a point source emission. The VHE $\gamma$-ray source G~0.9+0.1 \citep{Aharonian2005a} is located inside the MAGIC field-of-view. It shows a small excess consistent with the low flux reported by \citet{Aharonian2005a}. The MAGIC excess is not yet statistically significant for the given exposure time.

Figure \ref{fig:theta2} shows the distribution of the squared angular distance, $\theta^2$, between the reconstructed shower direction and the nominal GC position corresponding to Fig.~\ref{fig:prelresults} (observation periods II/III). The observed excess in the direction of the GC has a significance of 7.3 standard deviations ($\theta^{2} \leq 0.02^{\circ}$). The source position and the flux level are consistent with the measurement of HESS \citep{GC_hess} within errors.

For the determination of the energy spectrum, the Random Forest was trained including the source dependent image parameter DIST with respect to the nominal excess position. For the spectrum determination only the largest data set (period III) was used. The cut on the hadronness parameter (50\% $\gamma$-efficiency corresponding to an effective area of about 250000~m$^2$) resulted in about 500 excess events with a minimum SIZE of 200 ph.~el.
Figure \ref{fig:spectrum} shows the reconstructed VHE $\gamma$-ray energy spectrum of the GC after the unfolding with the instrumental energy resolution, see \citet{Mizobuchi2005}. The differential $\gamma$-flux can be well described by a simple power law:
\begin{eqnarray*}
\frac{\mathrm{d}N_{\gamma}}{(\mathrm{d}A \mathrm{d}t \mathrm{d}E)} = (2.9 \pm 0.6)\times 10^{-12} (E/\mathrm{TeV})^{-2.2 \pm 0.2} \nonumber
\\
\hspace{4cm} \mathrm{cm}^{-2}\mathrm{s}^{-1} \mathrm{TeV}^{-1} \ .
\end{eqnarray*}

The given errors ($1 \sigma$) are purely
statistical. The systematic error is estimated to be 35\%
in the flux level determination and 0.2 in the spectral index.

\begin{figure}[!h]
\begin{center}
\includegraphics[totalheight=6cm]{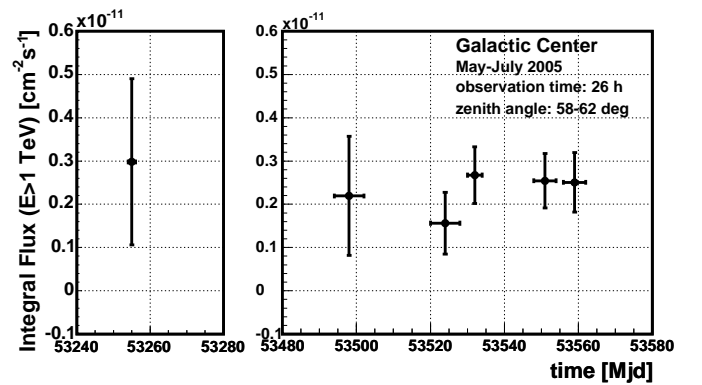}
\end{center}
\caption{Light curve: Reconstructed integral VHE gamma-ray flux above 1 TeV as a function of time. Within errors ($1 \sigma$) the data are consistent with a steady emission.}
\label{fig:light_curve}
\end{figure}

Figure \ref{fig:light_curve} shows the reconstructed integral VHE $\gamma$-ray flux above 1~TeV as a function of time. All OFF data are used for each time bin resulting in some correlation between the time bins. Different observation modi may result in different systematic errors. The flux level is steady within errors in the time-scales explored within these observations, as well as in the two year time-span between the MAGIC and HESS observations.

\vspace{-0.4cm}

\vspace{-0.1cm}

\section{Discussion}

Recent observations of TeV $\gamma$-rays from the GC confirm that
this is a very important region for high energy processes in the Galaxy.
In fact, this is not surprising since many different objects, 
able to accelerate particles above TeV energies, are expected there. The most likely source seems to be the massive black hole identified with
Sgr A$^*$ due to the directional consistency. 
A blazar-like relativistic jet originating from the spinning GC black hole
might be expected to produce TeV $\gamma$-rays \citep{Falcke1993}, but flux predictions of this model are on the low
side due to an unfavorable orientation of the jet axis. Moreover, a short-term
variability would be expected. 
\citet{Atoyan2004} propose that electrons can be accelerated to 
sufficiently high energies
at the termination shock of the sub-relativistic wind from the central part
of the advection dominated accretion flow onto the GC black hole, in analogy to the 
pulsar wind nebulae.  The authors explain the  broad band emission from
Sgr A$^*$ (from radio to TeV $\gamma$-rays) and suggest that the GeV source 
observed by EGRET has another origin. This is consistent with the recent determination of 
the position of the EGRET source 3EG J1746-2851 by \citet{Hooper2002} and \citet{Pohl2005}. Other scenarios for the $\gamma$-ray production in the vicinity of 
Sgr A$^*$, both leptonic and hadronic, have also been found to be consistent
with the TeV observations (for reasonable sets of parameters) but not
with the GeV observations \citep{Aharonian2005}.
It is generally expected that $\gamma$-rays produced in such compact source models
should show relatively fast variability. The same level of TeV flux reported 
by HESS in 2004 and by MAGIC in 2005, and also during their own observation 
periods extending over a few months, rather suggest a stable source on a year time 
scale. However, the $\gamma$-ray 
flux above 2.8 TeV ($3.7 \sigma$ significance) reported by Whipple during the extended period from 1995 through 2003 is a factor $\sim 2$ larger \citep{GC_whipple}.

The origin of $\gamma$-ray emission in other types of sources is also possible 
as demonstrated by the detection of the second TeV $\gamma$-ray source in the 
direction of the GC consistent with the location of the composite 
supernova remnant SNR G~0.9+0.1
\citep{Aharonian2005a}. \citet{Pohl1997} proposed that the GeV emission 
can be related to the GC radio arc. \citet{Crocker2005}, see also 
\citet{Fatuzzo2003}, argue for the GeV and TeV emission coming from different 
sites of the shell of the very powerful supernova remnant Sgr A East.

More extended $\gamma$-ray emission might also originate in the interaction of 
relativistic particles with the soft radiation and matter of the central 
stellar cluster around the GC.
These particles can be accelerated by e.g. a very energetic pulsar, 
a $\gamma$-ray burst source, shocks in the winds of the massive stars, or a 
shell type supernova remnant \citep{Bednarek2002,Biermann2004,Quataert2005,Crocker2005,Grasso2005}. 
If the TeV $\gamma$-rays are produced by leptons scattering off the infrared photons
from the dust heated by the UV stellar radiation
(as discussed by \citet{Quataert2005}), then  the $\gamma$-ray power at $\sim 100$ GeV
should be almost an order of magnitude higher due to scattering of UV 
radiation. The $\gamma$-ray energy spectrum should steepen between $\sim 0.1-1$ TeV.
Instead, the HESS collaboration reports a
simple power law spectrum between $\sim 0.2-10$~TeV \citep{GC_hess}. 
In order to produce a gamma-ray spectrum well described by a single
power law up to $\sim 20$~TeV, hadrons should have energies of about $10^{3}$~TeV. Such hadrons diffuse through the
region of the TeV source ($<7$ pc, \citet{GC_hess}) on a time scale of the order of
$10^4$ years, assuming the average magnetic field strength in this region
of $10^{-4}$~G and the Bohm diffusion coefficient.  
Therefore, the natural source of relativistic hadrons seems to be the supernova remnant
Sgr A East or the energetic pulsar created in the supernova explosion 
\citep{Crocker2005,LaRosa2005,Bednarek2002}. However, this relatively young source of relativistic 
hadrons cannot be identified with the last $\gamma$-ray burst 
in the center of our Galaxy if it appeared $\sim 10^6$ years ago \citep{Biermann2004}.

The GC can also be the brightest source of VHE $\gamma$-rays from particle dark matter annihilation \citep{Prada2004,Hooper2004,DM_MAGIC}. Most SUSY dark matter scenarios lead to a cut-off in the $\gamma$-ray energy spectrum below 10~TeV. The observed $\gamma$-ray energy spectrum extends up to 20~TeV. Thus most probably the main part of the observed $\gamma$-radiation is not due to dark matter annihilation \citep{Horns2004}. However, an
extended $\gamma$-ray source due to dark matter annihilation peaking in the region 10~GeV to 100~GeV
\citep{Elsaesser2005} cannot be ruled out yet.

\section{Concluding Remarks}

The MAGIC observations confirm the VHE $\gamma$-ray source at the Galactic
Center. The measured flux is compatible with the measurement of
HESS \citep{GC_hess} within errors. The VHE $\gamma$-ray emission does not show any
significant time variability; our measurements rather affirm a steady emission
of $\gamma$-rays from the GC region. The excess is point like, it's location is spatially consistent with SgrA$^*$ as well as SgrA East.

The nature of the source of the VHE $\gamma$-rays has not yet been identified. Future simultaneous observations 
with the present Cherenkov telescopes, the GLAST telescope and in the lower energies
will provide much better information on the source localization and variability of 
emission. This will shed new light on the nature of the high energy processes in the GC.

%------------------------------------------------------------------------------

%\appendix

\section*{Acknowledgements}

We would like to thank the IAC for the excellent working conditions at the Observatory de los Muchachos in La Palma. The support of the German BMBF and MPG, the Italian INFN and the Spanish CICYT is gratefully acknowledged. This work was also supported by ETH Research Grant TH-34/04-3 and the Polish MNiI Grant 1P03D01028.

%\newpage

%\bibliography{bibbib}

\begin{thebibliography}{}
\bibitem[Aharonian et~al.(2004)]{GC_hess}
Aharonian, F. et~al.,
\newblock 2004, A\&A, 425, L13.

\bibitem[Aharonian \& Neronov(2005)]{Aharonian2005}
Aharonian, F. \& Neronov, A.,
\newblock 2005, ApJ, 619, 306.

\bibitem[Aharonian et~al.(2005)]{Aharonian2005a}
Aharonian, F. et~al.,
\newblock 2005, A\&A, 432, L25.

\bibitem[Atoyan \& Dermer(2004)]{Atoyan2004}
Atoyan, A. \& Dermer, C.~D.,
\newblock 2004, ApJ, 617, L123.

\bibitem[Baixeras et~al.(2004)]{MAGIC-commissioning}
Baixeras, C. et~al. (MAGIC Collab.),
\newblock 2004, Nucl. Instrum. Meth., A518, 188.

\bibitem[Bednarek(2002)]{Bednarek2002}
Bednarek, W., 
\newblock 2002, MNRAS, 331, 483.

\bibitem[Biermann et~al.(2004)]{Biermann2004}
Biermann, P.~L. et al.,
\newblock  2004, ApJ, 604, L29.

\bibitem[Bock et~al.(2004)]{RF}
Bock, R.~K. et~al.,
\newblock 2004, Nucl. Instrum. Meth., A516, 511.

\bibitem[Bretz et~al.(2003)]{Bretz2003}
Bretz, T. et~al. (MAGIC Collab.),
\newblock 2003, Proc. of the 28th ICRC, Tsukuba, Japan, 2943.

\bibitem[Bretz \& Wagner(2003)]{Magic-software}
Bretz, T. \& R.~Wagner (MAGIC Collab.),
\newblock 2003, Proc. of the 28th ICRC, Tsukuba, Japan, 2947.

\bibitem[Cortina et~al.(2005)]{CortinaICRC}
Cortina, J. et~al. (MAGIC Collab.),
\newblock Proc. of the 29th ICRC, Pune, India, in press, astro-ph/0508274.

\bibitem[Crocker et~al.(2005)]{Crocker2005}
Crocker, R.M. et al., 
\newblock 2005, ApJ, 622, 892.

\bibitem[Domingo-Santamaria et~al.(2005)]{MAGIC_disp}
Domingo-Santamaria, E. et~al. (MAGIC Collab.),
\newblock  Proc. of the 29th ICRC, Pune, India, in press, astro-ph/0508274.

\bibitem[Els\"asser \& Mannheim(2005)]{Elsaesser2005}
Els\"asser, D. \& Mannheim, K., 
\newblock 2005, Phys. Rev. Lett., 94, 171302.

\bibitem[Falcke et~al.(1993) ]{Falcke1993}
Falcke, H., Mannheim, K., Biermann, P.L., 
\newblock 1993, A\&A, 278, L1.

\bibitem[Flix(2005)]{DM_MAGIC} 
Flix, J.,
\newblock 2005, Proc. of Rencontres de Moriond, La Thuile, Italy, astro-ph/0505313.

\bibitem[Fatuzzo \& Melia(2003)]{Fatuzzo2003}
M.~{Fatuzzo} and F.~{Melia},
\newblock 2003, ApJ, 596, 1035.

\bibitem[Fegan(1997)]{Fegan1997}
Fegan, D.~J.,
\newblock 1997, J. Phys. G., 23, 1013.

\bibitem[Fomin et~al.(1994)]{wobble}
Fomin, V.~P. et~al.
\newblock 1994, Astropart. Phys., 2, 137.

\bibitem[Grasso \& Maccione(2005)]{Grasso2005}
Grasso, D. \& Maccione, L.,
\newblock 2005, Astropart. Phys., 24, 273.

\bibitem[Gaug et~al.(2005)]{MAGIC_calibration}
Gaug, M. et~al. (MAGIC Collab.),
\newblock  Proc. of the 29th ICRC, Pune, India, in press, astro-ph/0508274.

\bibitem[Hillas(1985)]{Hillas_parameters}
Hillas, A.~M.,
\newblock 1985, Proc. of the 19th ICRC, La Jolla, USA. 

\bibitem[Hooper \& Dingus(2002)]{Hooper2002}
Hooper, D. \& Dingus, B.,
\newblock 2002, astro-ph/0212509.

\bibitem[Hooper et~al.(2004)]{Hooper2004}
Hooper, D. et~al.,
\newblock 2004, JCAPP, 9, 2.

\bibitem[Horns(2004)]{Horns2004}
Horns, D.,
\newblock 2005, Phys. Lett., B607, 225.

\bibitem[Kosack et~al.(2004)]{GC_whipple}
Kosack, K. et~al.,
\newblock 2004, ApJ, 608, L97.

\bibitem[LaRosa et~al.(2000)]{LaRosa2000}
LaRosa, T.~N. et al.,
\newblock 2000, AJ, 119, 207.

\bibitem[LaRosa et~al.(2005)]{LaRosa2005}
LaRosa, T.~N. et al.,
\newblock 2005, ApJ, 626, L23.

\bibitem[Lessard et~al.(2001)]{Lessard2001}
Lessard, R.~W. et~al.,
\newblock 2001, Astropart. Phys., 15, 1.

\bibitem[Mayer-Hasselwander et~al.(1998)]{GC_egret}
Mayer-Hasselwander, H.~A. et~al.,
\newblock 1998, A\&A, 335, 161.

\bibitem[Mizobuchi et~al.(2005)]{Mizobuchi2005}
Mizobuchi, S. et~al. (MAGIC Collab.),
\newblock Proc. of the 29th ICRC, Pune, India, in press, astro-ph/0508274.

\bibitem[Morris \& Serabyn(1996)]{GC_environment}
Morris, M. \& Serabyn, E.,
\newblock 1996, ARAA, 34, 645.

\bibitem[Pohl(1997)]{Pohl1997}
Pohl, M.,
\newblock 1997, A\&A, 317, 441.

\bibitem[Pohl(2005)]{Pohl2005}
Pohl, M.,
\newblock 2005, ApJ, 626, 174.

\bibitem[Prada et~al.(2004)]{Prada2004}
Prada, F. et~al.,
\newblock 2004, Phys. Rev. Lett., 93, 241301.

\bibitem[Quataert \& Loeb(2005)]{Quataert2005}
Quataert, E. and Loeb, A.
\newblock 2005, astro-ph/0509265. 

\bibitem[Riegel et~al.(2005)]{starguider}
Riegel, B. et~al. (MAGIC Collab.),
\newblock Proc. of the 29th ICRC, Pune, India, in press, astro-ph/0508274.

\bibitem[Rowell(2003)]{Rowell2003}
Rowell, G.~P.,
\newblock 2003, A\&A, 410, 398.

\bibitem[Sch\"odel et~al.(2002)]{Schoedel2002}
Sch\"odel, R. et~al.,
\newblock 2002, Nature, 419, 694.

\bibitem[Tsuchiya et~al.(2004)]{GC_cangaroo}
Tsuchiya, K. et~al.,
\newblock 2004, ApJ, 606, L115.

\bibitem[Wagner et~al.(2005)]{Crab_MAGIC}
Wagner, R. et~al. (MAGIC Collab.),
\newblock  Proc. of the 29th ICRC, Pune, India, in press, astro-ph/0508244.

\end{thebibliography}
%\bibliographystyle{GC}

\end{document}